\documentclass[a4paper,11pt]{article}

\usepackage[T1]{fontenc}
\usepackage[utf8]{inputenc}
\usepackage{lmodern}

\usepackage[a4paper,margin=1in,headheight=14pt]{geometry}
\usepackage{microtype}

\usepackage{array}
\usepackage{booktabs}
\usepackage{longtable}
\usepackage{tabularx}
\usepackage{ragged2e}
\usepackage{placeins}

\newcolumntype{L}{>{\RaggedRight\arraybackslash}X}
\newcolumntype{C}{>{\Centering\arraybackslash}X}
\newcolumntype{R}{>{\RaggedLeft\arraybackslash}X}
\newcolumntype{P}[1]{>{\RaggedRight\arraybackslash}p{#1}}

\usepackage{graphicx}
\graphicspath{{./}}

\usepackage[font=small,labelfont=bf]{caption}
\usepackage{subcaption} 

\usepackage[colorlinks=true,
            urlcolor=blue,
            citecolor=blue,
            linkcolor=black]{hyperref}
\title{Quantifying the Impact of CU: A Systematic Literature Review}

\author{Thomas Compton\\
University of York\\
\texttt{thomas.compton@york.ac.uk}}

\date{} 

\begin{document}

\maketitle
\section{abstract}
Community Unionism (CU) has served as a pivotal concept in debates on trade union renewal since the early 2000s, yet its theoretical coherence and political significance remain unresolved. This article investigates why CU has gained such prominence—not by testing its efficacy, but by mapping how it is constructed, cited, and contested across the scholarly literature. Using two complementary systematic approaches—a citation network analysis of 114 peer-reviewed documents and a thematic review of 18 core CU case studies—I examine how CU functions as both an empirical descriptor and a normative ideal. The analysis reveals CU’s dual genealogy: positioned by British scholars (e.g., Wills, 2002) as an ‘indigenous’ return to historic rank-and-file practices, yet structurally aligned with transnational social movement unionism. Thematic coding shows near-universal emphasis on coalition-building (100\%) and alliances (88\%), but deep ambivalence toward class politics (only 18\% explicitly embrace emancipatory aims). This tension suggests CU’s significance lies less in operationalising a new union model, and more in managing contradictions—between workplace and community, leadership and rank-and-file, reform and radicalism—within a shrinking labour movement.

\section{Introduction}
CU is a contested term. Wills and Simms (2004, p.61) argue ‘[T]he history of twentieth century community-union relations has evolved over time, moving from community-based trade unionism to representational community unionism.’ I begin with this quotation as it may have surprised may CU scholars who were arguing CU was primarily influenced by the American campaign ‘Justice for Janitors’. Moreover, the claim being made is not about the continuity in CU in Britain, that is assumed. Rather, it is that CU began as a more decentralised, or rank-and-file led part of unionism before becoming dominated by representatives. This is seen as a problem, with the article arguing that unions lost connection to local communities across the twentieth century and could now return to a form of ‘reciprocal’ CU, which would involve a community-union relationship where both benefits. However, the popular notion is that unions became ‘top-down’ or centralised organisations in the latter half of the nineteenth century (Fraser, 1999; Webb and Webb, 1894). This led to the rank-and-file movement, as covered by Hyman (1987). I begin with this point to suggest that CU literature is in a difficult position, deciding which traditions of union literature to situate itself within. Many CU scholars may view their approach as escaping some of the classic deadlocks of the industrial relations literature. What I wish to argue is that position is taken which would be familiar to the literature, but with the positions taken not given sufficient clarity, such that is it possible to identify whether CU has to be, for example, a rank-and-filist position.  

In beginning with this issue to highlight that the traditions of both unionists and academic researchers have been polarised into key positions. Workplace focused or community focused in one of these debates. However, using this framing, eschewing other substantial divides within the literature, may obscure more than it aids. I do not wish to argue which traditions are more academic and which are more representative of the attitudes of unionists. This may be better studied empirically. In this article, I wish to argue that these key tensions, community against workplace, rank-and-file against leadership, social movement against party politics imply a series of choices for scholar to take which are rarely made clear and kept consistent. These debates overlap with Hyman’s (2001) three union orientations, business (workplace), societal (social movement), class (rank-and-file), but not neatly. While some have sought to make these links, Hyman (2012) has argued their simplicity. I seek to go further and argue not only is Hyman (2001) correct to say unions often rely on multiple models, and this is useful for them, but that by reducing complex debates to ambiguous positions it has become difficult to accurately characterise each position. 

To explore this, I will begin with a debate on what community might mean and how it relates to the claims of CU. There is a tendency to claim communities have shared interests, highlighting how social movements have captured the interests of local communities. However, this is complicated by how broad the category social movement is, and whether it is fair to see unions as social movements. This issue could become the whole article, however, I wish to keep the debate within the literature of unions, rather than looking at social movement literature. This is where the article will become original, situating this debate in broader conflicts within the field. This will be pushed further in the second section, linking the CU approach to the tradition of respectability in unions. While Frege et al. (2004) and Hyman (2001) has made this tradition, the baggage this brings for CU scholars has not been teased out. As I shall highlight, once we look at broader debates on this tradition, it becomes less tenable to see it as a worthwhile project to return to. Just as, it becomes clear CU scholars are not seeking to emulate that tradition, picking and choosing parts of it which seem most appropriate. While this is a fair strategy, it becomes problematic when making claims as Wills and Simms (2004) do that historic forms of CU were more grassroots, and implicitly better than present forms. This is because their claim is premised on excluding the parts of union respectability that are unsavoury. Moreover, this causes consequences to the concept of CU, if it is premised on a historical form of unionism, should not there be a concern that there is a valorisation of historic practices which are were a problem for the unions. For example, misogyny.

\section{Trade union Models}

CU is one of many approaches to trade union strategy advocated and gained popularity in the early 2000s. Prominent alternatives include, 1) Marxist revivalism, 2) rank-and-file, 3) pluralism and 4) CU:
\begin{enumerate}
  \item Sheila Cohen (2006) provides an example of advocating a class-based approach and drawing on the longer history of Marxism in industrial relations (Hobsbawm, 1964; Hyman, 1975; Panitch, 1976; Thompson, 2016). This approach focuses on the ability of trade unions to act independently from middle-class institutions, working against class collaboration.
  \item Similar to Marxism in its class focus, rank-and-file approaches focus on industrial militancy and working-class action but take a position against the leadership of trade unions (Darlington, 2013; Darlington and Upchurch, 2012; Hyman, 1987; Zeitlin, 1989). Darlington (2009) is exemplary of this tradition, and Darlington and Upchurch (2012) draw explicitly on Marxism.
  \item Pluralism focuses on industrial collaboration where employers can collaborate with the employed for mutual benefit (Ackers, 2002; Dobbins et al., 2021; Kaufman and Gall, 2015). Ackers (2002) advocates for pluralism, continuing his focus on trade unions as employment-based organisations that both have no and should not engage themselves in equality-based activities (Ackers, 2001; 2015).
  \item CU focuses on common cause coalitions beyond the class basis of trade unions and focuses on equality (Fairbrother, 2008; Wills, 2002).
\end{enumerate}

\begin{figure}[htbp]
  \centering
  \includegraphics[width=0.9\textwidth]{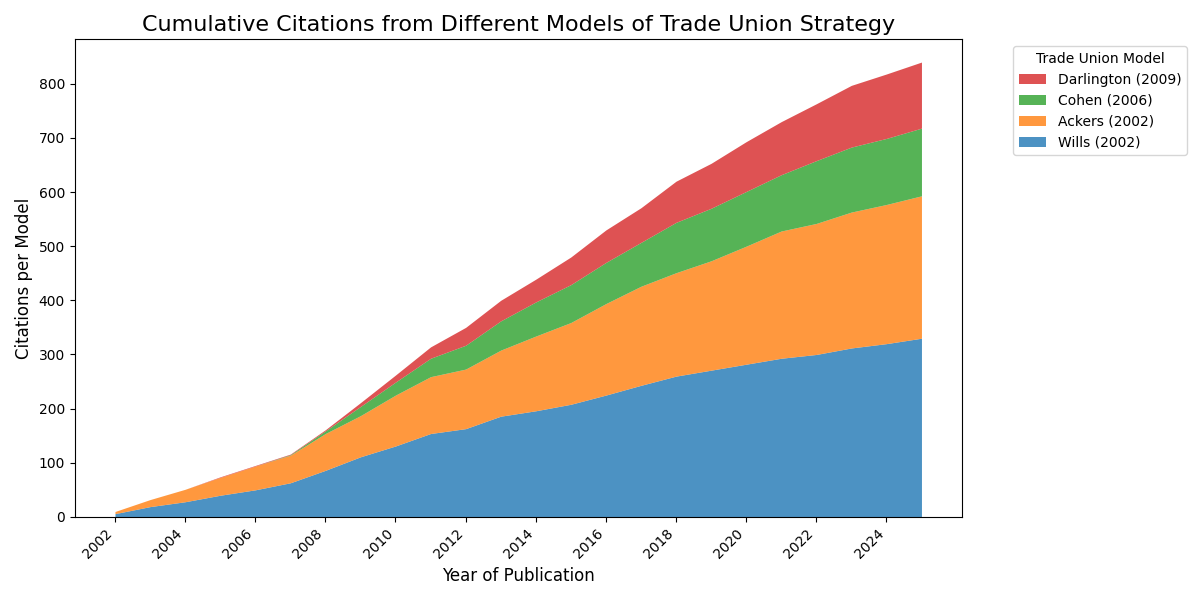}
  \caption{Cumulative citations from different models of trade union strategy (stacked area chart).}
\end{figure}

The more radical and political approaches of Marxism in Cohen (2006) and rank-and-filism in Darlington (2009) have garnered fewer citations than pluralism in Ackers (2002) and CU in Wills (2002) (see Figure 1). Wills (2002) saw its highest yearly citations in 2009 with 25, whereas Ackers (2002) peaked in 2010 with 18. As Figure 1 suggests, Wills (2002) has remained the most cited article of the four, with a 334 total citations and roughly 14 citations per year. CU has been deemed an appropriate approach for modern activities, as trade unions are no longer able to act in isolation because of their decreased memberships (Holgate, 2021; Lévesque and Murray, 2006). Especially with notions of community decline (Karn, 2007), it is unsurprising to see scholars view this approach as more relevant to modern social issues. Wills (2006, p. 913) summarises this position ‘[i]t is not sufficient to repeat the mantras of the past and fail to face up to the crisis [.]’ In Wills (2006) and Wills (2002) alike, Wills discusses the importance of coalitions between trade unions and different groups as the only appropriate solution to the current malaise of organised labour. As the citations of her article(s) reflect, this position has been adopted at least by the CU tradition and the trade union literature more broadly.

Once Wills (2002) had been established as a significant article, I used Publish or Perish to create a corpus of articles citing Wills (2002). The purpose of this corpus is to explore the wider impact of Wills (2002), quantifying direct citation (referred to as children) and citations of citations (grandchidlren). To find citations of citations, I selected 20 articles to have their citations quantified. The selection criteria focused on similarity to Wills (2002) in content and the number of citations. Articles such as Mollona (2009) and Alberti (2011) were kept despite having low citations because of their semantic similarity with Wills (2002). This means that all articles chosen as children were substantially engaged with Wills (2002) or utilised her framework. This selection process was undertaken manually.

\begin{figure}[htbp]
  \centering
  \includegraphics[width=0.8\textwidth]{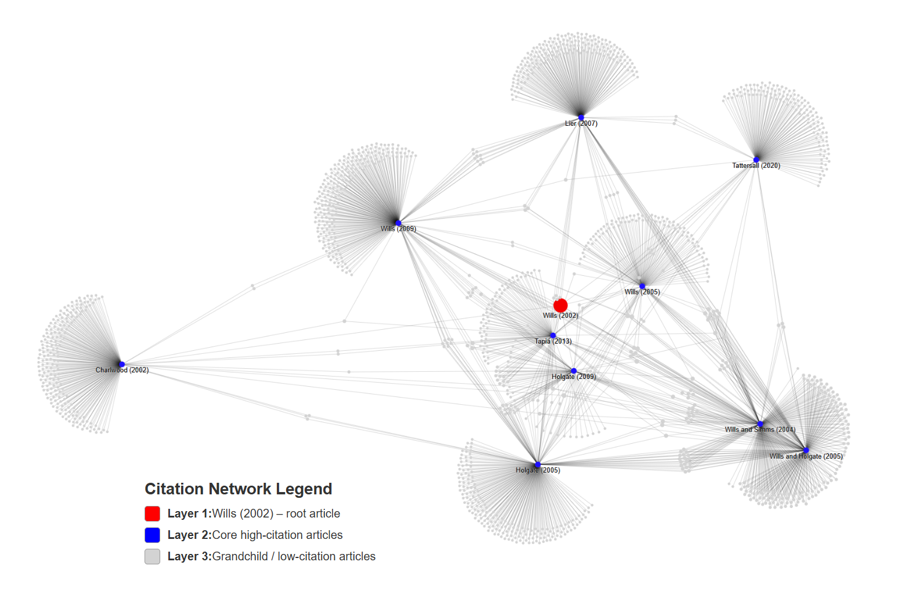}
  \caption{Citation Network  from Wills (2002) to 10 Highly-Cited CU Articles using Publish or Perish (Harzing, 2007), Gravis \& Network}
\end{figure}

\begin{table}[htbp]
\centering
\small
\caption{Influence counts by article}
\begin{tabularx}{\linewidth}{@{}X r r r@{}}
\toprule
\textbf{Article} & \textbf{Direct children} & \textbf{Grandchildren} & \textbf{Total influence} \\
\midrule
Wills (2002) & 317 & 371 & 688 \\
Wills and Simms (2004) & 226 & 222 & 448 \\
Wills and Holgate (2005) & 223 & 223 & 446 \\
Holgate (2005) & 264 & 84 & 348 \\
Wills (2009) & 272 & 56 & 328 \\
Lier (2007) & 213 & 43 & 256 \\
Charlwood (2002) & 210 & 15 & 225 \\
Wills (2005) & 127 & 46 & 173 \\
Tattersall (2020) & 144 & 23 & 167 \\
Holgate (2009) & 96 & 64 & 160 \\
Tapia (2013) & 101 & 46 & 147 \\
Holgate (2021) & 54 & 26 & 80 \\
Evans et al.\ (2007) & 48 & 18 & 66 \\
Alberti (2016) & 38 & 13 & 51 \\
Heery (2018) & 30 & 11 & 41 \\
Gall and Fiorito (2011) & 29 & 10 & 39 \\
Murray et al.\ (2013) & 31 & 3 & 34 \\
Savage and Wills (2004) & 21 & 11 & 32 \\
Heery et al.\ (2003) & 13 & 9 & 22 \\
Alberti (2011) & 3 & 0 & 3 \\
Mollona (2009) & 3 & 0 & 3 \\
\bottomrule
\end{tabularx}
\end{table}

While Figure 1 demonstrates that Wills (2002) has continued to be cited, Figure 2 demonstrates the broader impact of Wills (2002). Figure 2 shows the connections between Wills (2002), articles citing Wills (2002) and the connections between their citations. Therefore, Figure 2 supports the claim that Wills (2002) popularised CU in the British context, being both one of the earliest and most cited articles. Figure 2 also demonstrates the prominence of Wills and Holgate within the area, justifying the importance of exploring the claims of Holgate (2021) and Wills (2002). Wills (2002) defined the features of CU, with her article describing various trends within the labour movement evolving ‘beyond the fragments’ into CU. This article helped to introduce CU in Britain, while in the US John Sweeney popularised the term ‘social movement unionism’ (Lichtenstein, 2002; 2010). Wills (2002) and Wills and Simms (2004), focusing on British CU and arguing that this is a return to historic CU practices, appear to have popularised an idea of indigenous British trade union behaviours under the name CU, as opposed to the American Social Movement Unionism.  The concepts developed by Wills (2002) and Wills and Simms (2004) would be developed by other British case studies and literature reviews (Heery, 2018; Holgate, 2013; 2021; Holgate et al., 2021; Mollona, 2009). Therefore, this article has been successful in setting the agenda for CU in British articles. Fairbrother’s (2008) article, while less influential, summarises these contributions into a list of features, assisting in turning Wills (2002) discussion into features that can be validated.

This is evidence CU has attracted scholars who adopt the framework wholesale (e.g. Alberti, 2014; Holgate, 2013; Mollona, 2009), as well as those engaging more with the renewal tradition (e.g. Gall and Fiorito, 2011; Heery, 2018). Table 3 will explore a core corpus of CU articles, whereas Table 2 explores industrial relations articles relating to Wills (2002). This highlights the ambiguous status of some articles which discuss topics relating to CU but do not refer to themselves as CU articles, instead using the renewal framing, discussing opportunties for trade unions to advance. As shown in Figure 1, CU is one pathway of many available. This may explain why the framework is not always explicitly invoked where scholars seek to resolve the issue of renewal. At the same time, this may lead to confusion in what is CU, what the features of CU are and what the selection criteria for the core body of CU article should be.

Although Wills (2002) clearly established the features of CU, there are controversies regarding its applicability to different time periods.  Some scholars argue CU is a return to British traditions (Holgate, 2021; Mollona, 2009; Wills and Simms, 2004). Holgate (2021) links modern CU to historic friendly societies in their focus on self-help, where trade unionists are encouraged to resolve their issues instead of relying on trade union officials. The presumption is that there is a continuity between contemporary approaches to CU and historic variations from the twentieth and nineteenth centuries (Holgate, 2021; Wills and Simms, 2004). Len McClusky (2020; 2021), the former General Secretary of Unite, argues CU is a return to historic occupational communities, specifically in Liverpool. He focuses on how historic unions focused on communities surrounding the workplace (see also Allcorn and Marsh, 1975; Hobsbawm, 1989; Williams, 1962), as opposed to Wills’ (2002) focus on coalition building and going beyond the workplace.

\section{Systematic Review}

Using Publish or Perish, I created a corpus of articles on CU. I searched the term ‘Community Unionism’ to find relevant documents. This found 114 documents with at least one citation. After searching each document to ensure they are peer-reviewed articles and are primarily studying CU 18, articles remained. Each of these articles focuses on a case of CU, using primarily qualitative methods. The highest cited were Wills (2002) (342) and Wills and Simms (2004) (257), further emphasising their significance to the field. Their approach, alongside Holgate (2013), focuses on activism in London. Other articles discussed activity in Canada, America, Australia and South Africa. Therefore, CU appears to be most associated with anglophone countries, viewed as a cross-cultural phenomenon. This complicates the return to British traditions arguments, with cross-cultural influence being plausible. At the same time, the popularity of Wills (2002) suggests her article may have played a role in popularising this concept and allowing other scholars to explore its relevance to their countries. This is substantiated by the majority of articles proceeding Wills (2002).

\begin{table}[htbp]
\centering
\small
\caption{Summary of CU Corpora}
\label{tab:studies}
\begin{tabularx}{\linewidth}{@{}>{\RaggedRight\arraybackslash}p{3.2cm}
                              >{\RaggedRight\arraybackslash}p{2.8cm}
                              >{\RaggedRight\arraybackslash}X
                              >{\RaggedRight\arraybackslash}p{2.8cm}@{}}
\toprule
\textbf{Article} & \textbf{Methods} & \textbf{Subjects} & \textbf{Geographic focus} \\
\midrule
Holgate (2013) & interviews & London Citizens, charity workers & London (UK) \\
Wills (2002) & interviews & ISTC; Battersea \& Wandsworth TUC Organising Centre & London (UK) \\
Mollona (2009) & ethnography & Factory workers & Sheffield (UK) \\
Alberti (2014) & interviews & London Citizens & London (UK) \\
Wills \& Simms (2004) & interviews & Battersea \& Wandsworth TUC Organising Centre & London (UK) \\
Tufts (2002) & interviews & ILGWU & Toronto (CA) \\
Collins (2012) & secondary qualitative & n/a & Wisconsin (US) \\
Cranford \& Ladd (2003) & document analysis & Toronto Organising for Fair Employment & Toronto (CA) \\
Banks (1991) & secondary qualitative & SEIU & California (US) \\
Ellem (2003) & interviews & Pilbara Mineworkers Union & Pilbara (Aus) \\
Black (2005) & secondary qualitative & NMASS & New York (US) \\
Lucio \& Perrett (2009) & interviews & UNISON, ISTC, KFAT & North (UK) \\
Jordhus-Lier (2012) & secondary qualitative & SAMWU & Cape Town (SA) \\
Cockfield et al.\ (2008) & docs + interviews & AEU, MUA, ANF, CEPU, VIEU, CFMEU, TCFUA, AMWU, etc. & Victoria (Aus) \\
Atler (2013) & secondary qualitative & CTU & Chicago (US) \\
Pocock (2011) & mixed methods & n/a & Australia \\
Ross (2011) & interviews & CAW & Ontario (CA) \\
\bottomrule
\end{tabularx}
\end{table}
There are 6 articles studying the UK, 3 studying Canada (CA), 4 studying the United States (US), 1 studying South Africa (SA), and 3 studying Australia (AUS). Therefore, CU appears to be popular in the UK, with some articles also referring to Social Movement Unionism. This has been associated with America, despite CU being used in American contexts (see also Fine 1997; 2005). 

Discussing SMU, Fairbrother (2008, p. 214) summarises the four features as:
\begin{enumerate}
  \item locally focused and based, often referred to as rank-and-file mobilization, or variants thereof (Bezuidenhout 2000, 2; Turner 2003, 51).
  \item experimenting with collective actions that go beyond the strike or workplace-limited activities (Lambert 1998, 73).
  \item building alliances, coalition building, and extending into the community and beyond (Bezuidenhout 2000; Voss and Sherman 2000, 2003).
  \item embracing emancipatory politics, framing demands politically, and formulating transformative visions (Johnston 1994; Gindin 1995; Scipes 2003; Waterman 2001).
\end{enumerate}

\begin{table}[htbp]
\centering
\caption{Frequency of key themes in the 18-article CU corpus}
\label{tab:themes}
\begin{tabular}{@{}lrr@{}}
\toprule
\textbf{Theme} & \textbf{Occurrence} & \textbf{\%} \\
\midrule
coalition           & 17 / 17 & 100\% \\
alliance            & 15 / 17 & 88\%  \\
beyond the workplace & 11 / 17 & 65\%  \\
living wage         & 11 / 17 & 65\%  \\
equality            & 6 / 17  & 35\%  \\
rank-and-file       & 7 / 17  & 41\%  \\
reciprocal          & 7 / 17  & 41\%  \\
emancipatory        & 3 / 17  & 18\%  \\
\bottomrule
\end{tabular}
\end{table}

Of these features, the coalition or alliance building are well represented. Whereas, the expression 'beyond the workplace' has some popularity but is not universal. Equality or emancipatory goals are more controversial, with these more 'political' aims being less represented.

The living wage is more relevant for CU articles, as a trade union action that focuses on the use of public policy to achieve equality (Heery et al., 2019; Holgate, 2009; Wills, 2009). This is opposed to directly bargaining with employers for higher wages. However, this is still a focus on wages and not an attempt to rectify wage differentials.

\section{Which Community?}
There have been attempts at definitions of community unionism. Wills (2002, p.466) argues CU involves ‘common-cause alliances’ with ‘community groups’ and reaching ‘hard to organi[s]e groups’. This article is misleading by referring to CU as if it is an already existing concept, when what is being described has previously been referred to as ‘social movement unionism’, by John Sweeney (1996), its putative founder. I shall not delve into the disputed origins of CU, suffice it to say that it is viewed as synonymous with ‘social movement unionism’ and while there are conflicting definitions, what is outlined in Wills (2002) is reasonably consistent with the literature (Rainnie et al., 2010).

It is appropriate now to refer to specific definitions of Community Unionism. One definition arises from the social movement unionism literature, seeing the two as synonymous. This arises from the SEIU and then the AFL-CIO (Lichtenstein, 2002), as John Sweeney began at the former and moved to the latter. Key figures include John Sweeney (Sweeney, 1996), and Jane McAlevey (2003; 2016; 2020). It is difficult to highlight how many studies have begun by discussing 'Justice for Janitors' (see Clawson; 2003; Lichtenstein, 2002; 2010; Moody, 2022). This creates the issue that this is an American campaign, which may not be relevant to British unionism. Secondly, that it is a uniquely modern approach, which is why the case remains compelling. If it was what unions regularly do, then it would be difficult to see why it drew so much attention. Moreover, there is the issue of the difference between McAlevey (2003) and Sweeney (1996). For example, Sweeney (1996) brings a focus on electoral politics, supporting the Democratic party. Whereas, McAlevey (2003) is interested in bringing as many members as possible to union organising, which seems to focus on public policy campaigns. Both worked together reforming the AFL-CIO, with McAlevey (2003) building on the work of the SEIU (Clawson, 2003). Both would agree about increasing the memberships of the union, increasing the social standing of the union and reaching sympathetic groups. However, it is neither clear how successful both were with these campaigns and whether this was an issue of the concept advancing over time or going beyond what Sweeney (Sweeney, 1996) meant by social movement unionism. 

Increasing the number of active members would become associated with CU (see Holgate, 2021). This is where the idea of CU would overlap with the criticism of the servicing model. Sweeney would be associated with the usage of organisers, which would be imported into different countries interested in replicating Sweeney’s approach. This style of CU would appear more occupationally based, interested in developing the participation of members (Voss and Sherman, 2000). It is not necessary that unions have to chose between working with social movements or organising workplaces. Yet, organisers only have so much time, so unions may not have the resources to do both. Where this approach becomes controversial is in the usage of organisers as dictating union policy, instead of working with the rank-and-file to develop strategies related to their interests. Where the line must be drawn is between those who see CU as a rank-and-file project, or it is not CU, but rather the usage of organisers to create a managed activism. Put simply, how can it be community unionism if the community is not playing an important role (Wills and Simms, 2004), elsewise would it not be more appropriate to describe this as organising? 

On the other hand, there is the danger that this conception of CU is overly optimistic in what volunteers can achieve within the union. Thinkers such as Lenin, who was inspired by Webb and Webb (1920), saw the union executive as an important part of union representative democracy which was necessary for efficiency. Thus, it becomes an empirical issue that a rank-and-filist, CU may sound ideal, but whether it has happened and why not may need to be answered before the concept is taken as it currently stands. Again, this is what Cohen (2006) argues is the issue of prescriptivism, wherein scholars are creating concepts to describe solutions they would like to see as opposed to concepts which describe cases such as ‘Justice for Janitors’. Yet, this is always an issue when theorising union renewal strategies, which one could parse as normative and pragmatic concerns. For example, it may be good if unions can point to present social inequalities and attempt to resolve them. However, it is not clear whether unions have much ability to intervene on them, and whether this will provide advantage for the union. This is crucial if CU is a way of renewing the union, so a method of increasing the social standing of the union. For example, with the relationship the unions have with the Labour Party, it is very likely they could gain more political influences by supporting campaigns which increase social inequalities. Furthermore, there is an issue of union democracy not discussed, where it is a matter of what the memberships of the unions want out of their unions (Hyman, 2012). The members may not be so committed to social issues as academics, which may be an issue for union leaders who care about these issues, but not so much for ones who do not. 

This is why I have argued there may be a conflict between unions attempting to increase their diversity, and current members who would rather elect leaders who focus on workplace issues. I highlight these issues but make no claim that they are correct without empirical proof and are worthwhile to study. These are issues which have been raised to me in discussions with officers at unions, especially discussing Unite’s attempts to fight the far right with educational campaigns as concerns were raised when large amounts of their members were voting for the Conservative Party. Moreover, Sharon Graham, who would win the leadership election focused on workplace issues, against the explicit CU of Len McLusky. Yet, a more in depth case study would be required to fully understand the dynamics in the union and the broader social context. What has been certainly clear to Holgate et al. (2021) is that industrial members of Unite have been less than supportive of their community efforts. Therefore, CU is actually a democratic issue for the unions where they have to decide whether their constituents are the broader populations of society or whether it is primarily their memberships. This has been a long standing issue for unions (Fraser, 1999), which are often accused of being anti-social by elites (Hyman, 1989). 

Therefore, it could be fair to say that the CU may depend on how is included or excluded from the union’s conception of community. This is often the problem with defining a community, that a necessary exclusion is required, whether it is caused by lack of resources or prejudices. Whether the line is drawn between the political unionists and the more charitable focused, such as Sweeney against McAlevey. Whether it is a conflict between members focused on their wage concerns, and members interested in solving social issues as with Unite. Whether it is a conflict within the community between the leaderships and the rank-and-file, the issue emerges that the union, even when attempting to work with communities, must chose who it stands with and who it stands against. One of the issues not present in many community debates is that it may not be enough for unions to create a constituency, or a group with similar interests. Rather the issue is how to counter those in society who wish to end the union, and no degree of popular support will insulate them. As was the case where unions reached their peak membership in the same year Thatcher won a landslide victory, legislating many unions out of existence (Fraser, 1999; Wrigley, 2002). How far CU is a product of unions attempting to respond to hostile environments, or attempt to manage various tensions within and without the union could be useful to understand what encourages unions to undertake community activities, just as much as it is worth exploring what discourages them. 
\section{Which Union Tradition?}
So far, the key issue has been situating the CU movement, or scholastic tradition within the broader union context. I have argued there are key positions, such as rank-and-filism, or party entryism, or what Hyman would call political economism, which is the focus on wage bargaining. While it is certain that CU scholars are fighting against the latter position, it has been argued this may be a false dichotomy (Alberti, 2016; Mollona, 2009), especially as the same organisers may be working on public policy campaigns and organising workplaces. While there have been attempts to locate CU within Hyman’s (2001) societal orientation, the negative connotations of this position have not be thoroughly debated. Moreover, it should not be lost that Hyman (2001) argues that taking one orientation, between society, market and class, would not be ideal for any union, but rather the taking of a combination of them. This is contrary to the claims of CU as a strategy for renewing or revitalising the union, which seems to suggest that undertaking CU activities would increase the power of the unions. However, I do not wish to unfairly characterise this position. Different CU scholars have considered broader issues and solutions such as Holgate (2021). Yet, this creates the issue of whether CU remains limited to specific activities as outlined in different case studies, or whether a union undertaking these activities becomes a CU. For example, Heery (2002) discusses how the organising model being used in concert with the partnership approach may have undercut the efficacy of the strategy. Would a similar issue be relevant to CU, or rather is the issue of partnership a matter of CU being that it brings together different groups claiming they had shared interests. 

It is useful to begin with the case of Frege et al. (2004), which sought to argue that the miners’ strike was harmful to the community. Essentially, the miners should have put the community over their particular interests. This was a common anti-union argument historically, that unions in seeking to increase their pay packets were hurting the general public (Fraser, 1999). The issue is which community is impacted by this. Surely, the employers were the actors being anti-social, or anti-community in this situation, as we now understand the devasting community impacts of closing down the mines. What becomes an issue for the partnership, or cooperation model is that there are, at least, moments wherein the employers or the state will be acting against the unions and the broader community. CU scholars appear happy to argue the social impacts of austerity, but these are thoroughly political battles where in different communities, or peoples are created and fought against. Rather than there being a generalised community with unifying interests, these moments of mobilisation create different groups on either side, in what we may call as contentious politics. What may be surprising is that many case studies have been interested in this side of contentious politics, seeing how the London Citizens assembly has been used to contradict politicians. However, doing this creates similar democratic issues for the unions wherein they are choosing sides within the community and criticising the political system which is accepted by most of the public. 
This was a similar issue facing unions who supported cooperation or respectability historically. For examples, Raynes (1921) discusses the history of a union which began by claiming it would not strike, but eventually was pushed by its members to undertake industrial action. Unions have the rank-and-file to contend with which is a community which may have concerns at odds with the broader community. This is an issue Hobsbawm (1989) discusses in regard to public sector unions in which the industrial actions of one could have trade implications for the other. Unions may be able to suppress the wage demands of workers (Mills, 2001), this may appeal to other groups within society, but also create an environment where union leaders receive state rewards for suppressing the legitimate grievances of their workers (Hyman, 1989). Again, this is choosing one community over the other, and would be argued as benefiting society. However, it may be helpful here to draw a distinction between society and community. Society may mean the total national community, which is at some level fictious, whereas community refers more practical to a local entity. While a strike would still, potentially, negatively impact community and society, it could be helpful to see the local community related to the union as more on the side of their activities. In part, this is what CU scholars seek to argue, that unions maybe be disliked by the elites of a society, they may have better luck with the members of their local communities who have either shared interests or connections to the union which could be leveraged to gain support for the union’s endeavours. 

One may wonder why the term class would not be more appropriate. Moreover, why claiming communities have shared interests is any less suspect than claiming the working class has shared interests. Of course, this is a matter of which tradition to draw on, where there have been extensive cases of writers arguing that cross-class alliances are better than unions remaining within the working class. While this argument is not as common in modern times, the argument that unions should remain class-focused is equally uncommon. For example, Hyman (2001) does not argue this position, despite situating himself within the Marxist tradition. In some way, this can be useful because arguing unions should focus on class approaches may seem overly prescriptive. However, arguing unions should not remain class-focused can be, equally, perspective. Yet, the issue may be more practical, such that unions with some level of class support need some coherent strategy to effectively use this. Just as unions, which may be able to gain some elite-level support, may not be able to effectively utilise their support to achieve their goals, and may have to give up too much influence, such that the collaboration is not in their advantage. These are often difficult claims to make in the abstract. With the example of the Labour Party, it may be fairly easy to argue the lack of advantage given by this party, given the various historical cases of the party acting against the interests of the union through the support of anti-union legislation. This situation has a clear negative outcome for union, whereas campaigns such as living wage support have less clear advantages for the unions. They may increase the conditions for workers, but essentially create new free riders, but also worse, they give workers higher wages not secured through collective bargaining, therefore giving them less reason to join the unions. 

The question of tradition, therefore, is difficult when situating the CU scholarship. This is an historical issue, insofar as it is a debate on which historical cases should be called CU, but also which history of unions is drawn upon. I have argued that CU scholarship may wish to present their approach as an innovation from the 1990s, but also bring back old arguments from cross-class alliance scholars, or what would be called the liberal tradition. Yet, at the same time, CU is argued as a renewal strategy, which presents it as a radical break from the type of liberalism present in most unions. Sometimes radical social movements are referred to as sources of inspiration. This is not developed sufficiently, and writers such as Holgate (2021) appear in some situations to be supporting the anti-capitalism of Occupy, and the pro-capitalism of the Labour Party. This create confusion in defining the political program for CU and defining whether being on the side of community means being on the side of capitalism. Just as there are many historical cases of unions being accused of being anti-social, it becomes an issue of whose community, where terms like class may be helpful to delineate between elites happy to end ways of living by closing jobs. Fundamentally, it appears difficult for unions to please everyone, especially given there are elites within most societies which influence popular opinion and wish to criminalise, or at least stigmatise, unions. Finally, there is the question of why scholars need to take a position of being on the side of class or community. It appears that there is some overlap between the two and that the scholarly debate tends to be between whether the union should focus on collective bargaining or social issue campaigning. While scholars have been taking positions, it is important to ask why do this when it is possible to support both at the same time. This is where the issue of renewal becomes important, because if the author is offering a way of fixing the issues of the unions, it would require that their approach is unique, and not what unions are currently undertaking, such that if the proposal were undertaken, it would be possible to delineating a break where the ‘turn’ began. Actually proving this is quite difficult empirically, given the heterogeneity within and between unions, such that it may not be helpful to characterise a large number of unions. 

\section{Conclusion}
While the issue of union renewal was important in the 1990s and early 2000s, it appears as though this literature has decreased in popularity. There has been a decrease in empirical work in studying union attempts to increase their social standing. However, in terms of the decline of proposals for union renewal, the record is more mixed. The problems for offering solutions to union decline begins with how complex unions are and the reasons for their decline are. Unions themselves are complex organisations which may have organisers who work both on increasing memberships and running campaigns to influence public policy. Equally, unions are declining for various reasons including the long-term effects of anti-union policies, changes in business approaches to labour retention, and the general decline in not-for-profits. These issues alone could merit their own article. However, some scholars have argued social movements offer a new model for unions. Yet, modern movements have been quick mobilisations which quickly dissipated. Unions have outlasted various social movements. While many movements have radical politics which could be imported into unions, this would be controversial among communities. It has been argued that using social movement strategies and public support for them could help unions, but many modern social movements have been highly controversial and would involve alienating sections of the community, potentially, just as much as they would garner support for the unions. Moreover, the collaboration with social movements creates issues of democracy within unions where scholars may be asked which democracy should be taking priority, wherein union democracies, however, limited, may reflect the opinions of the members, which may be part of what hinders the union’s radical approach to politics. 

Furthermore, the question of union politics is ambiguous within the CU literature. In Britain, the large unions remain supportive of a Labour Party which has delivered some positives but has done very little to assist with anti-union legislation. Moreover, the party has shown to be resistant to influence from social movements and unions alike, suggesting that any union becoming more social movement-focused would risk further conflicting with the Labour Party. This creates a concern about whether unions are able to effectively influence public policy. This would require empirical investigation to understand how far unions have been effective in their bargaining but would still require a political verdict on how far union agitation of policymakers is a worthwhile activity or risks becoming more focused on performing radicality rather than using their limited resources efficiently. 

\section{References}
\begin{itemize}
    \item Ackers, P. (2002). Reframing employment relations: The case for neo-pluralism. Industrial relations journal, 33 (1), pp.2–19. 
    \item Ackers, P. (2015). Trade unions as professional associations. Finding a voice at work, pp.95–126. 
    \item Alberti, G. (2016). Moving beyond the dichotomy of workplace and community unionism: The challenges of organising migrant workers in London’s hotels. Economic and Industrial Democracy, 37 (1), pp.73–94. 
    \item Allcorn, D. H. and Marsh, C. M. (1975). Occupational Communities - Communities of What? In: Bulmer, M. (Ed). Working-Class Images of Society (Routledge Revivals). Routledge. pp.206–218.
    \item Clawson, D. (2003). The Next Upsurge: Labor and the New Social Movements. Ithaca, NY: ILR Press. 
    \item Cohen, S. (2006). Ramparts of resistance: Why workers lost their power and how to get it back. Pluto Press. 
    \item Darlington, R. (2009a). Organising, Militancy and Revitalisation: The Case of the RMT Union. In: Gall, G. (Ed). Union Revitalisation in Advanced Economies: Assessing the Contribution of Union Organising. London: Palgrave Macmillan UK. pp.83–106. 
    \item Darlington, R. (2009b). RMT strike activity on London underground: incidence, dynamics and causes. In: International Labour Process Conference, Edinburgh. 2009. pp.6–8.
    \item Darlington, R. (2018). The leadership component of Kelly’s mobilisation theory: Contribution, tensions, limitations and further development. Economic and Industrial Democracy, 39 (4), pp.617–638. [Online]. Available at: doi:10.1177/0143831X18777609.
    \item Darlington, R. and Upchurch, M. (2012). A reappraisal of the rank-and-file versus bureaucracy debate. Capital \& Class, 36 (1), pp.77–95.
    \item Fairbrother, P. (2008). Social movement unionism or trade unions as social movements. Employee Responsibilities and Rights Journal, 20, pp.213–220.
    \item Fine, J. (1997). Community unionism: The key to the new labor movement. Perspectives on Work, pp.32–35.
    \item Fraser, W. H. (1999). A History of British Trade Unionism 1700–1998. Bloomsbury Publishing.
    \item Frege, C. M., Heery, E. and Turner, L. (2004). The new solidarity? Trade union coalition-building in five countries. Varieties of unionism: Strategies for union revitalization in a globalizing economy, p.137.
    \item Harzing, A. W. (2007). Publish or Perish. [Online]. Available at: \\ https://harzing.com/resources/publish-or-perish.
    \item Heery, E. (2002). Partnership versus organising: alternative futures for British trade unionism. Industrial Relations Journal, 33 (1), pp.20–35. 
    \item Heery, E. (2009). Trade unions and contingent labour: scale and method. Cambridge Journal of Regions, Economy and Society, 2 (3), pp.429–442.
    \item Heery, E. (2018). Fusion or replacement? Labour and the ‘new’ social movements. Economic and Industrial Democracy, 39 (4), pp.661–680.
    \item Heery, E., Hann, D. and Nash, D. (2019). Going it alone? The involvement of trade unions in the Living Wage campaign in the United Kingdom. Employment Relations in the 21st Century: Challenges for Theory and Research in a Changing World of Work. Alphen aan den Rijin: Kluwer Law International BV, pp.105–122.
    \item Hobsbawm, E. J. (1989). Politics for a rational left: political writing, 1977-1988. Verso. 
    \item Holgate, J. (2009). Contested terrain: London’s living wage campaign and the tensions between community and union organising. In: McBride, J. and Greenwood, I. (Eds). Community Unionism: A Comparative Analysis of Concepts and Contexts. Springer. pp.49–74.
    \item Holgate, J. (2021). Arise: Power, Strategy and Union Resurgence. London: Pluto Press.
    \item Hyman, R. (1987). Rank-and-file movements and workplace organisation, 1914-39. In: Wrigley, C. (Ed). A History of British Industrial Relations. 2. Harvester Press. pp.129–158.
    \item Hyman, R. (2001). Understanding European trade unionism: between market, class and society. Sage.
    \item Hyman, R. (2012). Will the real Richard Hyman please stand up? Capital \& Class, 36 (1), pp.151–164.
    \item Lévesque, C. and Murray, G. (2006). How do unions renew? Paths to union renewal. Labor Studies Journal, 31 (3), pp.1–13.
    \item Lichtenstein, N. (2002). A race between cynicism and hope: labor and academia. New Labor Forum, pp.71–79.
    \item Lichtenstein, N. (2010). Why American Unions Need Intellectuals. Dissent, 57 (2), pp.69–73.
    \item McAlevey, J. (2003). It Takes a Community: Building Unions from the Outside In. New Labor Forum, 12 (1), pp.22–32.
    \item McAlevey, J. (2016). No shortcuts: Organizing for power in the new gilded age. Oxford University Press.
    \item McAlevey, J. (2020). A Collective Bargain: Unions, Organizing, and the Fight for Democracy. New York, NY: Ecco.
    \item McCluskey, L. (2020). Why You Should Be a Trade Unionist. London New York: Verso.
    \item McCluskey, L. (2021). Always Red. OR Books.
    \item Mills, C. W. (2001). The new men of power: America’s labor leaders. University of Illinois Press. 
    \item Mollona, M. (2009). Community unionism versus business unionism: The return of the moral economy in trade union studies. American Ethnologist, 36 (4), pp.651–666.
    \item Rainnie, A., McGrath-Champ, S. and Herod, A. (2010). Making space for geography in labour process theory. In: Thompson, P. and \item Smith, C. (Eds). Working life: renewing labour process analysis. Critical perspectives on work and employment series. Basingstoke: Palgrave Macmillan. pp.297–315.
    \item Sweeney, J. J. (1996). America Needs a Raise: Fighting for Economic Security and Social Justice. Boston: Houghton Mifflin.
    \item Webb, S. and Webb, B. (1894). The history of trade unionism. Longmans, Green.
    \item Williams, J. E. (1962). The Derbyshire Miners: A Study in Industrial and Social History. First Edition. George Allen and Unwin Ltd.
    \item Wills, J. (2002). Community unionism and trade union renewal in the UK: moving beyond the fragments at last? Transactions of the Institute of British Geographers, 26 (4), pp.465–483. [Online]. Available at: doi:10.1111/1475-5661.00035.
    \item Wills, J. (2009). The living wage. Soundings, 42 (1), pp.33–46.
    \item Wills, J. and Simms, M. (2004). Building reciprocal community unionism in the UK. Capital \& Class, 28 (1), pp.59–84. [Online]. Available at: doi:10.1177/030981680408200105.
    \item Wrigley, C. (2002). British trade unions since 1933. Cambridge University Press.
\end{itemize}

\end{document}